# Mathematical Analyses

# of an accelerating (Griffiths-Podolsky) Black Hole

Kartheek R Solipuram

Chaitanya Bharathi Institute of Technology, Gandipet, Hyderabad, India. 500075

**Abstract**

An exact solution of Einstein's equations which represents a pair of accelerating and rotating black holes was presented by J. B. Griffiths and J. Podolsky [2]. In the paper [2] they have shown the explicit form of a spinning C-metric starting from the Plebanski-Demianski metric, and transformed it using NUT and angular velocity parameters in addition to the usual parameters and thus gave a generalized form of such solutions. In the forthcoming discussion, an attempt has been made to realize the Riemann components of the proposed metric. Furthermore, certain optical characteristics of the metric have been analyzed using the Newman-Penrose formalism.

## I. Introduction

The well known black hole metrics are the Schwarzchild and Kerr metrics which elucidate the fields around static and rotating black holes respectively. Their charged versions, Reissner- Nordstrom and Kerr-Newman solutions are also known; so is the spinning C-metric. A large family of electrovacuum solutions was proposed by Plebanski and Demianski [1] to which some additional parameters like the NUT parameter and the angular velocity parameter have been included to give a convenient form:

$$ds^2 = \frac{1}{\Omega^2}\{\frac{Q}{\rho^2}[dt-(a\sin^2\theta + 4l\sin^2\theta/2)]^2 - \frac{\rho^2}{Q}dr^2 - \frac{\tilde{P}}{\rho^2}[adt-(r^2+(a+l)^2)d\phi^2]^2 - \frac{\rho^2}{\tilde{P}}\sin^2\theta d\theta^2\}$$

(1)

where

$$\Omega = 1 - \frac{\alpha}{\omega}(l+a\cos\theta)r$$

$$\rho^2 = r^2 + (l+a\cos\theta)^2$$

$$\tilde{P} = \sin^2\theta(1 - a_3\cos\theta - a_4\cos^2\theta)$$

$$Q = [(\omega^2 k + e^2 + g^2)(1+2\frac{\alpha l}{\omega}r) - 2mr + \frac{\omega^2 k}{a^2 - l^2}r^2] \times [1+\frac{\alpha(a-l)}{\omega}r][1-\frac{\alpha(a+l)}{\omega}r]$$

(1a)

And

$$a_3 = 2\frac{\alpha a}{\omega}m - 4\frac{\alpha^2 al}{\omega^2}(\omega^2 k + e^2 + g^2)$$

$$a_4 = -\frac{\alpha^2 a^2}{\omega^2}(\omega^2 k + e^2 + g^2)$$

(1b)

With $k$ given as,

$$(\frac{\omega^2}{a^2-l^2} + 3\alpha^2 l^2)k = 1 + 2\frac{\alpha l}{\omega}m - 3\frac{\alpha^2 l^2}{\omega^2}(e^2+g^2).$$

Here, the various parameters have been obtained by various transformations, inclusions and the consideration of the requisite canonical forms and after all this we have $m$ as the mass parameter, $l$ as the NUT parameter, $e$ as the electric parameter, $g$ as the magnetic parameter, $\alpha$ as the acceleration parameter, and $\omega$ as the angular velocity parameter. Thus, we have seven arbitrary parameters $m, l, e, g, a, \alpha, \omega$ in all. The above metric may be called G-P metric for future references.

**II. Requisite Treatment**

The treatment in the forthcoming sections will be very similar to the second chapter in [3]. From the various simplifications, and the usual conventions, let us assume the general form of the metric to be

$$ds^2 = e^{2\upsilon}(dt - \Gamma d\phi)^2 - e^{2\psi}(dt - \Upsilon d\phi)^2 - e^{2\mu_2}(dx^2)^2 - e^{2\mu_3}(dx^3)^2 \qquad (2)$$

Here, the parameter $\Gamma$ is similar to $\omega$ in the usual stationary axisymmetric metric. The new parameter $\Upsilon$ however has been added for the generalization of the G-P metric. We conserve the signature $(+1, -1, -1, -1)$ preserving it to be of a local Minkowski space time. We proceed to find the Riemann components by considering the Cartan's equations of structure.

For this let us take,

$$ds^2 = e^{2\upsilon}(dt - \Gamma d\phi)^2 - e^{2\psi}(dt - \Upsilon d\phi)^2 - \sum_A e^{2\mu_A}(dx^A)^2 \ ; \ A = 2, 3, \qquad (2a)$$

where the Latin indices $(A, B, ....)$ and the summations over them are restricted to the values 2, 3.

In applying Cartan's method [3] we shall take,

$$\omega^A = e^{\mu_A} dx^A$$
$$\omega^1 = e^{\psi}(dx^0 - \Upsilon dx^1). \qquad (3)$$
$$\omega^0 = e^{\upsilon}(dx^0 - \Gamma dx^1)$$

From these we can readily get the inverse relations,

$$dx^A = e^{-\mu_A}\omega^A$$

$$dx^1 = \frac{e^{-\psi}\omega^1 - e^{-\upsilon}\omega^0}{\Upsilon - \Gamma} \qquad (3a)$$

$$dx^0 = \frac{e^{-\upsilon}\omega^0\Upsilon - e^{-\psi}\omega^1\Gamma}{\Upsilon - \Gamma}$$

The above forms retain the $(+,-,-,-)$ signature for the so chosen orthonormal tetrad.

Now, it follows that,

$$d\omega^A = \sum_B e^{-\mu_B}\mu_{A,B}\omega^B \wedge \omega^A + \frac{1}{\Upsilon - \Gamma}[(\mu_{A,1} - \Gamma\mu_{A,0})e^{-\psi}\omega^1 \wedge \omega^A + (\Upsilon\mu_{A,0} - \mu_{A,1})e^{-\upsilon}\omega^0 \wedge \omega^A]$$

$$d\omega^1 = \psi_{,1}[\frac{1}{\Gamma - \Upsilon}e^{-\upsilon}\omega^1 \wedge \omega^0 - \frac{2\Upsilon}{(\Gamma - \Upsilon)^2}e^{\psi - 2\upsilon}\omega^0 \wedge \omega^0 - \frac{\Gamma + \Upsilon}{(\Gamma - \Upsilon)^2}e^{-\upsilon}\omega^1 \wedge \omega^1]$$

$$d\omega^0 = \upsilon_{,0}[\frac{\Gamma^2 + \Gamma\Upsilon}{(\Gamma - \Upsilon)^2}e^{-\psi}\omega^1 \wedge \omega^0 - \frac{2\Gamma^2}{(\Gamma - \Upsilon)^2}e^{\upsilon - 2\psi}\omega^1 \wedge \omega^1 - \frac{\Upsilon^2 - \Gamma\Upsilon}{(\Gamma - \Upsilon)^2}e^{-\upsilon}\omega^0 \wedge \omega^0]$$

(4)

This is the expression of $\omega$'s in terms of the exterior derivatives $\omega^i \wedge \omega^j$,

$(i \neq j, i, j = 1,2,3,4)$.

Cartan's first equations of structure suggest [torsion is zero]:

$$d\omega^0 = -\sum_A \omega^0_A \wedge \omega^A - \omega^0_1 \wedge \omega^1$$

$$d\omega^1 = -\sum_A \omega^1_A \wedge \omega^A - \omega^1_0 \wedge \omega^0 \qquad , \qquad (5)$$

$$d\omega^A = -\sum_B \omega^A_B \wedge \omega^B - \omega^A_1 \wedge \omega^1 - \omega^A_0 \wedge \omega^0$$

following which we have the results,

$$\omega_2^1 = \frac{1}{\Upsilon - \Gamma}(\mu_{2,1} - \Gamma\mu_{2,0})e^{-\psi}\omega^2 = Xe^{-\psi}\omega^2$$

$$\omega_3^1 = \frac{1}{\Upsilon - \Gamma}(\mu_{3,1} - \Gamma\mu_{3,0})e^{-\psi}\omega^3 = Ae^{-\psi}\omega^3$$

$$\omega_2^0 = \frac{1}{\Upsilon - \Gamma}(\Upsilon\mu_{2,0} - \mu_{2,1})e^{-\upsilon}\omega^2 = Ze^{-\upsilon}\omega^2$$

$$\omega_3^0 = \frac{1}{\Upsilon - \Gamma}(\Upsilon\mu_{3,0} - \mu_{3,1})e^{-\upsilon}\omega^3 = Ye^{-\upsilon}\omega^3$$

$$\omega_3^2 = e^{-\mu_3}\mu_{2,3}\omega^2 + e^{-\mu_2}\mu_{3,2}\omega^3$$

$$\omega_1^0 = \frac{1}{\Gamma - \Upsilon}\psi_{,1}e^{-\upsilon}\omega^1 - \frac{2\Upsilon}{(\Gamma - \Upsilon)^2}\psi_{,1}e^{\psi - 2\upsilon}\omega^0 - \frac{\Gamma^2 + \Gamma\Upsilon}{(\Gamma - \Upsilon)^2}\upsilon_{,0}e^{-\psi}\omega^0 + \frac{2\Upsilon^2}{(\Gamma - \Upsilon)^2}\upsilon_{,0}e^{\upsilon - 2\psi}\omega^1$$

$$= F_{1'}\psi_{,1}e^{-\upsilon}\omega^1 - F_{2'}\psi_{,1}e^{\psi - 2\upsilon}\omega^0 - F_{3'}\upsilon_{,0}e^{-\psi}\omega^0 + F_{4'}\upsilon_{,0}e^{\upsilon - 2\psi}\omega^1$$

$$= F_1\omega^1 - F_2\omega^2 - F_3\omega^0 + F_4\omega^1$$

(5a)

Here, the various symbols, $X, A, Z, Y, F_{1'}, F_{2'}, F_{3'}, F_{4'}, F_1, F_2, F_3, F_4$ that can be easily followed from the foregoing expressions have been used are for future purposes to ease the length of expressions.

Finally, to calculate the Riemann components, we introduce the following lemmas which are readily verifiable.

LEMMAS:

If $\Pi$ is a function of $dx^0, dx^1, dx^2, dx^3$, then, we have:

$$d(\Pi\omega^1) = \frac{(\Pi\psi)_{,1}}{(\Gamma-\Upsilon)^2}[(\Gamma-\Upsilon)e^{-\upsilon}\omega^1\wedge\omega^0 - 2\Upsilon e^{\psi-2\upsilon}\omega^0\wedge\omega^0 - (\Gamma+\Upsilon)e^{-\psi}\omega^1\wedge\omega^1]$$

$$d(\Pi\omega^0) = \frac{(\Pi\upsilon)_{,0}}{(\Gamma-\Upsilon)^2}[(\Gamma^2+\Gamma\Upsilon)e^{-\psi}\omega^1\wedge\omega^0 - 2\Gamma^2 e^{\upsilon-2\psi}\omega^1\wedge\omega^1 - (\Gamma^2-\Gamma\Upsilon)e^{-\upsilon}\omega^0\wedge\omega^0]$$

$$d(\Pi\omega^A) = e^{-\mu_B}(\Pi\mu_A)_{,B}\omega^B\wedge\omega^A + \frac{1}{\Gamma-\Upsilon}\{[(\Gamma(\Pi\mu_A)_{,0}-(\Pi\mu_A)_{,1})e^{-\psi}\omega^1\wedge\omega^A]+$$

$$[((\Pi\mu_A)_{,1}+\Upsilon(\Pi\mu_A)_{,0})e^{-\upsilon}\omega^0\wedge\omega^A]\}$$

(6)

Thus, having procured all the necessary grounds we can proceed further toward the calculation of the Riemann components for the proposed general form of the G-P metric.

### III. Riemann Components

We make use of the Cartan's second equation of structure [3], namely,

$$\frac{1}{2}R^i_{jkl}\omega^k\wedge\omega^l = \Omega^i_j = d\omega^i_j + \omega^i_k\wedge\omega^k_j, \qquad (7)$$

For which the necessary exterior derivatives have been evaluated in the previous discussion. The lemmas have been designed for the purpose of evaluating the Riemann components.

Thus it is clear that $R^i_{jkl}$ is the coefficient of $\omega^k\wedge\omega^l$ in the expression on the right in the above equation.

Now, $R^0_{1kl}$ is the coefficient of $\omega^k\wedge\omega^l$ in the corresponding equation of structure,

$$\frac{1}{2}R^0_{1kl}\omega^k\wedge\omega^l = \Omega^0_1 = d\omega^0_1 + \omega^0_k\wedge\omega^k_1$$, and following from the lemmas we have,

$$d\omega_1^0 = \frac{1}{2}\{[\frac{((F_1+F_4)\psi)_{,1}}{(\Gamma-\Upsilon)^2}((\Gamma-\Upsilon)e^{-\upsilon}\omega^1\wedge\omega^0 - 2\Upsilon e^{\psi-2\upsilon}\omega^0\wedge\omega^0 - (\Gamma+\Upsilon)e^{-\psi}\omega^1\wedge\omega^1)] -$$
$$[\frac{((F_2+F_3)\upsilon)_{,0}}{(\Gamma-\Upsilon)^2}((\Gamma^2+\Gamma\Upsilon)e^{-\psi}\omega^1\wedge\omega^0 - 2\Gamma^2 e^{\upsilon-2\psi}\omega^1\wedge\omega^1 - (\Upsilon^2-\Gamma\Upsilon)e^{-\upsilon}\omega^0\wedge\omega^0)]\}$$

(7a)

And so, $R_{0101}$ (the coefficient of $\omega^0 \wedge \omega^1$ in $\Omega_1^0$) is,

$$R_{0101} = \frac{1}{2}\{[\frac{((F_2+F_3)\upsilon)_{,0}}{(\Gamma-\Upsilon)^2}e^{-\psi}(\Gamma^2+\Gamma\Upsilon)] - [\frac{((F_1+F_4)\psi)_{,1}}{\Gamma-\Upsilon}e^{-\upsilon}]\}. \tag{8a}$$

Accordingly, the Riemann components $R_{0102}, R_{0103}, R_{0113}, R_{0123}, R_{0112}$ are found to vanish.

Thus,

$$R_{0102} = R_{0103} = R_{0113} = R_{0123} = R_{0112} = 0.$$

Similarly we establish $R_{2kl}^0$ which is the coefficient of $\omega^k \wedge \omega^l$ in the equation,

$$\frac{1}{2}R_{2kl}^0 \omega^k \wedge \omega^l = \Omega_2^0 = d\omega_2^0 + \omega_k^0 \wedge \omega_2^k,$$ and the required exterior derivative is,

$$d\omega_2^0 = (Ze^{-\upsilon}\mu_2)_{,B} e^{-\mu_B}\omega^B \wedge \omega^2 + \frac{1}{\Gamma-\Upsilon}\{[(\Gamma(Ze^{-\upsilon}\mu_A)_{,0} - (Ze^{-\upsilon}\mu_A)_{,1})e^{-\psi}\omega^1\wedge\omega^2] +$$
$$[((Ze^{-\upsilon}\mu_A)_{,1} + \Upsilon(Ze^{-\upsilon}\mu_A)_{,0})e^{-\upsilon}\omega^0\wedge\omega^2]\}$$

(7b)

Hence, $R_{0202}$ (the coefficient of $\omega^0 \wedge \omega^2$ in $\Omega_2^0$) is,

$$R_{0202} = \frac{1}{\Gamma-\Upsilon}[((Ze^{-\upsilon}\mu_2)_{,1} + \Upsilon(Ze^{-\upsilon}\mu_2)_{,0})e^{-\upsilon}] - F_2 X e^{-\psi} - F_3 X e^{-\psi}, \tag{8b}$$

$R_{0212}$ (the coefficient of $\omega^1 \wedge \omega^2$ in $\Omega_2^0$) is,

$$R_{0212} = \frac{1}{\Gamma-\Upsilon}[((Xe^{-\psi}\mu_2)_{,1} + \Upsilon(Xe^{-\psi}\mu_2)_{,0})e^{-\upsilon}] + F_2 Z e^{-\upsilon}\psi_{,1} e^{\psi-2\upsilon} + F_3 Z e^{-\upsilon}\upsilon_{,0} e^{-\psi} \tag{8c}$$

$R_{0223}$ (the coefficient of $\omega^2 \wedge \omega^3$ in $\Omega_2^0$) is,

$$R_{0223} = -e^{\mu_3}(Xe^{-\upsilon}\mu_2)_{,3} - e^{-\mu_3}\mu_{2,3} Y e^{-\upsilon}. \tag{8d}$$

In this case the Riemann components that vanish are $R_{0203}, R_{0213}$.

Proceeding in this fashion, we have,

$$R_{0303} = \frac{1}{\Gamma - \Upsilon}[((Ye^{-\upsilon}\mu_3)_{,1} + \Upsilon(Ye^{-\upsilon}\mu_3)_{,0})e^{-\upsilon}] - F_2 Ae^{-\psi} - F_3 Ae^{-\psi}, \tag{8e}$$

$$R_{0313} = \frac{1}{\Gamma - \Upsilon}[(\Gamma(Ye^{-\upsilon}\mu_3)_{,0} - (Ye^{-\upsilon}\mu_3)_{,1})e^{-\psi}] + F_1 Ae^{-\psi} + F_4 Ae^{-\psi}, \tag{8f}$$

$$R_{0323} = e^{-\mu_3}(Ye^{-\upsilon}\mu_3)_{,3} + Ze^{-\upsilon}e^{-\mu_2}\mu_{3,2}. \tag{8g}$$

$$R_{0312} = 0. \tag{8h}$$

$$R_{1223} = e^{-\mu_3}(Xe^{-\psi}\mu_2)_{,3} + Ae^{-\psi}e^{-\mu_3}\mu_{2,3}, \tag{8i}$$

$$R_{1212} = \frac{1}{\Gamma - \Upsilon}[\Gamma(Xe^{-\psi}\mu_2)_{,0} - (Xe^{-\psi}\mu_2)_{,1}] - F_1 e^{-\upsilon}Ze^{-\upsilon} - F_4 e^{\upsilon - 2\psi}Ze^{-\upsilon}. \tag{8j}$$

$$R_{1213} = 0 \tag{8k}$$

$$R_{1313} = \frac{1}{\Gamma - \Upsilon}[(\Gamma(Ae^{-\psi}\mu_3)_{,0} - (Ae^{-\psi}\mu_3)_{,1})e^{-\psi}] - F_1 Ye^{-\upsilon} - F_4 Ye^{-\upsilon}, \tag{8l}$$

$$R_{1323} = e^{-\mu_2}(Ae^{-\psi}\mu_3)_{,2} + Xe^{-\psi}e^{-\mu_2}\mu_{3,2}. \tag{8m}$$

$$R_{2323} = e^{-\mu_2}(e^{-\mu_2}\mu_{3,2}\mu_3)_{,2} - e^{-\mu_3}(e^{-\mu_3}\mu_{2,3}\mu_2)_{,3} - ZYe^{-2\upsilon}. \tag{8n}$$

So, in the case of the G-P metric we have 12 non-vanishing Riemann components. And, these components include the electric, magnetic and NUT parameters as well.

**IV. Null Tetrad**

One of the most important tools in deciphering the physical interpretation of any black hole metric is the Newman Penrose (N-P) formalism. In this context we shall analyze the particular characteristics by evaluating the spin coefficients.

For the suggested G-P metric, we ought to list the covariant and contravariant basis before we go any further. So, we shall consider its geodesic in the equatorial plane. Now, on the equatorial plane, we have $\theta = \pi/2 =$ constant, and $\dot{\theta} = 0$. Hence the Lagrangian becomes,

$$2L = \frac{1}{\Omega^2}\{[\frac{Q-a^2}{r^2+l^2}\dot{t}^2 + (2Q\frac{a+2l}{r^2+l^2} + \frac{2a}{r^2+l^2}(r^2+(a+l)^2)) - \frac{r^2+l^2}{Q}\dot{r}^2 + (\frac{Q}{r^2+l^2}(a+2l)^2 - \frac{(r+(a+l)^2)^2}{r^2+l^2})\dot{\phi}^2]\} \tag{9}$$

where $\Omega = 1 - \frac{\alpha l r}{\omega}$

And we deduce the generalized momenta to be,

$$p_t = \frac{1}{\Omega^2}[\frac{Q-a^2}{r^2+l^2}\dot{t} + (\frac{a+2l}{r^2+l^2}Q + \frac{2a}{r^2+l^2}(r^2+(a+l)^2))\dot{\phi}] = \text{I}$$

$$-p_\phi = \frac{1}{\Omega^2}[-(\frac{a+2l}{r^2+l^2}Q + \frac{2a}{r^2+l^2}(r^2+(a+l)^2))\dot{t} - (\frac{Q}{r^2+l^2}(a+2l)^2 - \frac{(r^2+(a+l)^2)^2}{r^2+l^2})\dot{\phi}] = \text{K}$$

$$-p_r = \frac{r^2+l^2}{Q}\dot{r}$$

(10)

The constancy of $p_t$ and $p_\phi$ follows from the independence of the Lagrangian on $t$ and $\phi$, which is in fact a manifestation of the stationary and axisymmetric character of the chosen metric. The superior dots used in the above set of equations denote differentiation with respect to an affine parameter $\tau$.

The Hamiltonian is given by

$$\text{H} = p_t\dot{t} + p_\phi\dot{\phi} + p_r\dot{r} - L$$
$$= \frac{1}{2}\frac{Q-a^2}{r^2+l^2}\dot{t}^2 + (\frac{a+2l}{r^2+l^2}Q + \frac{2a}{r^2+l^2}(r^2+(a+l)^2))\dot{t}\dot{\phi} - \frac{r^2+l^2}{2Q}\dot{r}^2 +, \tag{11}$$
$$\frac{1}{2}(\frac{(a+2l)^2}{r^2+l^2}Q - \frac{(r^2+(a+l)^2)^2}{r^2+l^2})\dot{\phi}^2$$

And from the independence of the Hamiltonian on $t$ we get,

$$2H = \frac{1}{\Omega^2}[\frac{Q-a^2}{r^2+l^2}\dot{t} + (\frac{a+2l}{r^2+l^2}Q + \frac{2a}{r^2+l^2}(r^2+(a+l)^2)^2)\dot{\phi}]\dot{t}$$
$$-[(\frac{a+2l}{r^2+l^2}Q + \frac{2a}{r^2+l^2}(r^2+(a+l)^2)^2)\dot{t} - (\frac{Q}{r^2+l^2} - \frac{(r^2+(a+l)^2)^2}{r^2+l^2})\dot{\phi}]\dot{\phi} \quad , \tag{12}$$
$$= I\dot{t} - K\dot{\phi} - \frac{r^2+l^2}{Q}\dot{r}^2 = \delta_1$$

where $\delta_1$ is a constant. We may set $\delta_1 = 0$ for null geodesics. The above equation can thus be simplified for null geodesics to get,

$$(r^2+l^2)\dot{r}^2 = Q(I\dot{t} - K\dot{\phi}). \tag{13}$$

The equations for the constants I and K can be treated as simultaneous linear equations in $\dot{t}$ and $\dot{\phi}$ of the form,

$$\Omega^2 I = P\dot{t} + N\dot{\phi}$$
$$\Omega^2 K = O\dot{t} + T\dot{\phi} \tag{14}$$

where,
$$P = \frac{Q-a^2}{r^2+l^2},$$
$$N = \frac{a+2l}{r^2+l^2}Q + \frac{2a}{r^2+l^2}(r^2+(a+l)^2),$$
$$O = \frac{a+2l}{r^2+l^2}Q + \frac{2a}{r^2+l^2}(r^2+(a+l)^2), \tag{14a}$$
$$T = \frac{(a+2l)^2}{r^2+l^2}Q - \frac{(r^2+(a+l)^2)^2}{r^2+l^2}$$

Solving these simultaneous equations, we have,
$$\dot{\phi} = \frac{O\Omega^2 I - P\Omega^2 K}{NO - PT},$$
$$\dot{t} = \frac{\Omega^2 P(IT + KN)}{NO - PT} \tag{15}$$

Rewriting the above lines by calling upon an impact parameter $D = K/I$ by putting

$D = a$ and $K = aI$ as

$$\dot{\phi} = I\Omega^2 \frac{O - aP}{NO - PT} = I\Omega^2 \varpi,$$
$$\dot{t} = I\Omega^2 \frac{PT + aPN}{NO - PT} = I\Omega^2 \vartheta$$
(15a)

where $\varpi$ and $\vartheta$ have their designated meanings.

Substituting these values into form of null geodesics we had procured before, we get,

$$\dot{r} = \pm I\Omega\sqrt{Q\varpi - Qa\vartheta}.$$
(16)

From the above calculations we get the general form of geodesics as,

$$\frac{\partial t}{\partial \tau} = I\Omega^2 \vartheta; \frac{\partial r}{\partial \tau} = \pm I\Omega\sqrt{Q\varpi - Qa\vartheta}; \frac{\partial \theta}{\partial \tau} = 0; \frac{\partial \phi}{\partial \tau} = I\Omega^2 \varpi.$$
(17)

Now, to delineate the null tetrad using the conventional N-P vectors, $l, n, m, \bar{m}$ using all the necessary orthogonalization and normalization conditions:

$$l \cdot m = l \cdot \bar{m} = n \cdot m = n \cdot \bar{m} = 0,$$
$$l \cdot n = 1,$$
$$m \cdot \bar{m} = -1$$
(18)

So, the G-P metric has the null tetrad,

$$l^i = \frac{1}{\Lambda}(\Omega \upsilon \quad \Lambda \quad 0 \quad \Omega \varpi)$$
$$n^i = \frac{1}{2\rho^2}(\Omega \upsilon \quad -\Lambda \quad 0 \quad \Omega \varpi)$$
$$m^i = \frac{1}{\bar{\rho}\sqrt{2}}(i(l+a)\sin\theta \quad 0 \quad 1 \quad i\csc\theta)$$
(19a)

The corresponding covariant tetrad is,

$$l_i = \frac{1}{\Lambda}(\Lambda \quad \Omega\vartheta \quad -\Omega\varpi \quad 0)$$
$$n_i = \frac{1}{2\rho^2}(-\Lambda \quad \Omega\vartheta \quad -\Omega\varpi \quad 0)$$
$$m_i = \frac{1}{\bar{\rho}\sqrt{2}}(0 \quad i(l+a)\sin\theta \quad -i\csc\theta \quad -1)$$
(19b)

In each case $\bar{m}_i$ can be obtained by considering the complex conjugates of the components in $m_i$.

Moreover also note that,

$$\Lambda = \Omega I \sqrt{Q\varpi - Qa\vartheta} = \Omega I \Delta,$$
$$\Delta = \sqrt{Q\varpi - Qa\vartheta}$$
$$\rho^2 = r^2 + (l+a)^2 \cos^2\theta, \quad . \tag{20}$$
$$\bar{\rho} = r + i(l+a)\cos\theta,$$
$$\bar{\rho}^* = r - i(l+a)\cos\theta$$

Here, $\bar{\rho}^*$ is the complex conjugate of $\bar{\rho}$.

**V. Spin Coefficients**

The Ricci rotation coefficients in the tetrad formalism are actually the spin coefficients in the N-P formalism. The various spin coefficients are,

$$\kappa = \gamma_{311} = \frac{1}{2}(\lambda_{311} + \lambda_{131} - \lambda_{113})$$
$$\lambda_{311} = \lambda_{131} = 0,$$
$$\kappa = 0.$$

$$\sigma = \gamma_{313} = \frac{1}{2}(\lambda_{313} + \lambda_{331} - \lambda_{133})$$
$$\lambda_{313} = \lambda_{331} = 0,$$
$$\sigma = 0.$$

$$\lambda = \gamma_{244} = \frac{1}{2}(\lambda_{244} + \lambda_{424} - \lambda_{442})$$
$$\lambda_{244} = \lambda_{424} = 0,$$
$$\lambda = 0.$$

$$\nu = \gamma_{242} = \frac{1}{2}(\lambda_{242} + \lambda_{224} - \lambda_{422})$$
$$\lambda_{242} = \lambda_{224} = 0,$$
$$\nu = 0. \qquad . \tag{21}$$

These vanishing coefficients suggest that the congruence of null geodesics, $l$ and $n$ are shear-free. And from the shear-free character of these congruences we can conclude on the basis of Goldberg-Sachs [4] theorem, that the G-P space time is of Petrov [5] type D. Continuing, with the rest of the coefficients,

$$\rho = \gamma_{314} = \frac{1}{2}(\lambda_{314} + \lambda_{431} - \lambda_{143})$$

$$\lambda_{314} = 0$$

$$\lambda_{431} = \frac{((l+a)\sin\theta)^2 - i(l+a)\bar{\rho}\cos\theta}{2\bar{\rho}^2\bar{\rho}^*}(\varpi + \vartheta)\Omega$$

$$-\lambda_{143} = \frac{i(l+a)\sin\theta}{2\Lambda\bar{\rho}(\bar{\rho}^*)^2}\Omega\varpi + \frac{i(l+a)\bar{\rho}^*\cos\theta + ((l+a)\sin\theta)^2}{2\bar{\rho}(\bar{\rho}^*)^2} - \frac{i\csc\theta}{2\bar{\rho}(\bar{\rho}^*)^2},$$

$$2\rho = \frac{((l+a)\sin\theta)^2 - i(l+a)\bar{\rho}\cos\theta}{2\bar{\rho}^2\bar{\rho}^*}(\varpi + \vartheta)\Omega +$$

$$\frac{i(l+a)\sin\theta}{2\Lambda\bar{\rho}(\bar{\rho}^*)^2}\Omega\varpi + \frac{i(l+a)\bar{\rho}^*\cos\theta + ((l+a)\sin\theta)^2}{2\bar{\rho}(\bar{\rho}^*)^2} - \frac{i\csc\theta}{2\bar{\rho}(\bar{\rho}^*)^2}$$

$$\mu = \gamma_{243} = \frac{1}{2}(\lambda_{243} + \lambda_{324} - \lambda_{432})$$

$$\lambda_{243} = \frac{l+a-i\bar{\rho}^*\csc\theta\cot\theta}{4\rho^2\bar{\rho}(\bar{\rho}^{*2})}\Lambda + \frac{i\csc\theta}{4\rho^2\bar{\rho}(\bar{\rho}^{*2})}\Lambda + \frac{i(l+a)\sin\theta}{4\rho^2\bar{\rho}(\bar{\rho}^{*2})}\Omega\varpi$$

$$\lambda_{324} = 0$$

$$-\lambda_{432} = \frac{i(l+a)\bar{\rho}\cos\theta + ((l+a)\sin\theta)^2}{4\rho^2\bar{\rho}^*(\bar{\rho})^2} + \frac{2i\csc\theta}{4\rho^2\bar{\rho}^*(\bar{\rho})^2}\Lambda + \frac{i(l+a)\sin\theta}{4\rho^2\bar{\rho}^*(\bar{\rho})^2}\Omega\varpi \qquad (22)$$

$$2\mu = \frac{l+a-i\bar{\rho}^*\csc\theta\cot\theta}{4\rho^2\bar{\rho}(\bar{\rho}^{*2})}\Lambda + \frac{i\csc\theta}{4\rho^2\bar{\rho}(\bar{\rho}^{*2})}\Lambda +$$

$$\frac{i(l+a)\sin\theta}{4\rho^2\bar{\rho}(\bar{\rho}^{*2})}\Omega\varpi + \frac{i(l+a)\bar{\rho}\cos\theta + ((l+a)\sin\theta)^2}{4\rho^2\bar{\rho}^*(\bar{\rho})^2} + \frac{2i\csc\theta}{4\rho^2\bar{\rho}^*(\bar{\rho})^2}\Lambda + \frac{i(l+a)\sin\theta}{4\rho^2\bar{\rho}^*(\bar{\rho})^2}\Omega\varpi$$

Before we proceed any further, we must make some more calculations which will prove worthwhile in the forthcoming coefficients. Particularly, we are concerned with a few partial derivatives.

Firstly let us denote, for any function, $W$ of $x^0, x^1, x^2, x^3$,

$$\frac{\partial}{\partial r} W = W_r .\tag{23}$$

$$\frac{\partial}{\partial r}(\Omega \varpi) = (\Omega \varpi)_r = \Omega_r \varpi + \Omega \varpi_r$$

$$\Omega_r = -\frac{\alpha l}{\omega}$$

$$\varpi_r = \frac{1}{f^2}\{[2r(2lQ + 2ar^2 + M) + (r^2 + l^2)(2Q_r l + 4ar)]f - [(2lQ + 2ar^2 + M)(r^2 + l^2)]f_r\}$$

$$Q_r = \{[(\omega^2 k + e^2 + g^2)(\frac{2\alpha l}{\omega}) - 2m + \frac{2\omega^2 kr}{a^2 - l^2}][1 + \frac{\alpha(a-l)}{\omega}r][1 - \frac{\alpha(a+l)}{\omega}r]$$

$$+[(\omega^2 k + e^2 + g^2)(1 + \frac{2\alpha lr}{\omega}) - 2mr + \frac{\omega^2 kr^2}{a^2 - l^2}][1 + \frac{\alpha(a-l)}{\omega}][1 - \frac{\alpha(a+l)}{\omega}r] +$$

$$[(\omega^2 k + e^2 + g^2)(1 + \frac{2\alpha lr}{\omega}) - 2mr + \frac{\omega^2 kr^2}{a^2 - l^2}][1 + \frac{\alpha(a-l)}{\omega}r][1 - \frac{\alpha(a+l)}{\omega}]\}$$

$$f_r = \{Q_r[Q(a+2l)^2 - (r^2 + (a+l)^2)^2] + [(Q - a^2)(Q_r - 4r^3 - 4r(a+l)^2)] - [Q_r(a+2l) + 4ar]\}$$

$$\vartheta_r = \frac{1}{f^2}\{[Q_r(Q(a+2l) + 2a(r^2 + (a+l)^2)) + (Q - a^2)(Q_r + 4ar) - Q_r(Q(a+2l)^2$$

$$-(r^2 + (a+l)^2)^2) - (Q - a^2)(Q_r - 4r^3 - 4ar^2)]f + f_r[(Q - a^2)(Q(a+2l) + 2a(r^2 + (a+l)^2))$$

$$-(Q - a^2)(Q(a+2l)^2 - (r^2 + (a+l)^2))]\}$$

$$\Lambda_r = \frac{1}{2}\Delta^{3/2}[Q_r(\varpi - a\vartheta) + Q(\varpi_r - a\vartheta_r)]\Omega I + \Delta\Omega_r I$$

$$.\tag{24}$$

In the foregoing set of expressions, we have taken,

$$\varpi = \frac{(2lQ + 2ar^2 + M^2)(r^2 + l^2)}{f},$$

$$\vartheta = \frac{1}{f}\{[(Q - a^2)(Q(a+2l) + 2a(r^2 + (a+l)^2))] -,\tag{25}$$

$$[(Q - a^2)(Q(a+2l)^2 - (r^2 + (a+l)^2)^2)]\}$$

where,

$$M^2 = 3a^2 + 2al^2 + 2a^2l$$
$$f = [(Q-a^2)(Q(a+2l)^2 - (r^2+(a+l)^2)^2)] - .$$
$$[Q(a+2l) + 2a(r^2+(a+l)^2)]$$
(25a)

We now continue to evaluate the remaining coefficients.

$$\tau = \gamma_{312} = \frac{1}{2}(\lambda_{312} + \lambda_{231} - \lambda_{123})$$

$$\lambda_{312} = \frac{1}{2\sqrt{2}\bar{\rho}\rho^2}[(\Omega\varpi)_r - \frac{\Omega\varpi}{\Lambda}\Lambda_r]$$

$$\lambda_{231} = \frac{i\csc\theta}{\sqrt{2}(\rho\bar{\rho})^2}$$

$$-\lambda_{123} = \frac{(l+a)^2\sin\theta\cos\theta}{\sqrt{2}\rho^4\bar{\rho}}\Omega\vartheta - \frac{\rho^2\Lambda_r - 2r\Lambda}{2\sqrt{2}\rho^4\bar{\rho}}(i(l+a)\sin\theta) - \frac{\Omega\varpi}{2\sqrt{2}\rho^4\bar{\rho}}$$

$$2\tau = \frac{1}{2\sqrt{2}\bar{\rho}\rho^2}[(\Omega\varpi)_r - \frac{\Omega\varpi}{\Lambda}\Lambda_r] + \frac{i\csc\theta}{\sqrt{2}(\rho\bar{\rho})^2} + \frac{(l+a)^2\sin\theta\cos\theta}{\sqrt{2}\rho^4\bar{\rho}}\Omega\vartheta -$$

$$\frac{\rho^2\Lambda_r - 2r\Lambda}{2\sqrt{2}\rho^4\bar{\rho}}(i(l+a)\sin\theta) - \frac{\Omega\varpi}{2\sqrt{2}\rho^4\bar{\rho}}$$

$$\pi = \gamma_{241} = \frac{1}{2}(\lambda_{241} + \lambda_{124} - \lambda_{412})$$

$$\lambda_{241} = -\frac{i\csc\theta}{\sqrt{2}(\rho\bar{\rho}^*)^2}\Omega\varpi$$

$$\lambda_{124} = \frac{1}{2\sqrt{2}\rho^4\bar{\rho}^*}\{[(\Omega\varpi)_r\rho^2 - 2r\Omega\varpi] - i(l+a)\sin\theta(\Lambda_r\rho^2 - 2r\Lambda)\}$$

$$-\lambda_{412} = \frac{1}{2\sqrt{2}\rho^2\bar{\rho}^*\Lambda^2}[(\Omega\varpi)_r\Lambda - \Omega\varpi\Lambda_r]$$

$$2\pi = -\frac{i\csc\theta}{\sqrt{2}(\rho\bar{\rho}^*)^2}\Omega\varpi + \frac{1}{2\sqrt{2}\rho^4\bar{\rho}^*}\{[(\Omega\varpi)_r\rho^2 - 2r\Omega\varpi] -$$

$$i(l+a)\sin\theta(\Lambda_r\rho^2 - 2r\Lambda)\} + \frac{1}{2\sqrt{2}\rho^2\bar{\rho}^*\Lambda^2}[(\Omega\varpi)_r\Lambda - \Omega\varpi\Lambda_r]$$

(26)

$$\varepsilon = \frac{1}{2}(\gamma_{211} + \gamma_{341})$$

$$\gamma_{211} = \frac{1}{2}(\lambda_{211} + \lambda_{121} - \lambda_{112})$$

$$\lambda_{121} = \lambda_{211} = 0$$

$$\gamma_{211} = 0. \tag{27}$$

$$\gamma_{341} = \frac{1}{2}(\lambda_{341} + \lambda_{134} - \lambda_{413})$$

$$\lambda_{413} = 0$$

$$\lambda_{341} = \frac{1}{2\bar{\rho}(\bar{\rho}^*)^2}[(i(l+a)\sin\theta)^2 - i\bar{\rho}^*(l+a)\sin\theta] + \frac{i\csc\theta}{\bar{\rho}(\bar{\rho}^*)^2} - \frac{i(l+a)\sin\theta}{2\Lambda\bar{\rho}(\bar{\rho}^*)^2}$$

In the above evaluation, we see that $\lambda_{341}$ is the complex conjugate of $\lambda_{134}$, and therefore, we can write,

$$\lambda_{341} = (R^1 + R^2 + R^3) + i(C^1 + C^2 + C^3), \tag{28}$$

where,

$$\Sigma R^1 = -r\Lambda(l+a)^2 \sin^2\theta(1+\cos\theta)$$

$$\Sigma R^2 = -2\Lambda(l+a)\csc\theta\cos\theta$$

$$\Sigma R^3 = (l+a)^2 \sin\theta\cos\theta$$

$$\Sigma C^1 = -(l+a)\Lambda\sin\theta[(l+a)^2 \sin\theta + r + (l+a)^2 \cos^2\theta]. \tag{28a}$$

$$\Sigma C^2 = 2r\Lambda\csc\theta$$

$$\Sigma C^3 = -r(l+a)\sin\theta$$

$$\Sigma = 2(r^2 + (l+a)^2 \cos^2\theta)^2$$

Thus we have,

$$2\varepsilon = i(C^1 + C^2 + C^3). \tag{29}$$

$$\gamma = \frac{1}{2}(\gamma_{212} + \gamma_{342})$$

$$\gamma_{212} = \frac{1}{2}(\lambda_{212} + \lambda_{221} - \lambda_{122})$$

$$\lambda_{212} = 0$$

$$\lambda_{221} = -\frac{\Omega \vartheta}{2\rho^6}(\rho^2 \Lambda_r - 2r\Lambda)$$

$$\gamma_{212} = -\frac{\Omega \vartheta}{2\rho^6}(\rho^2 \Lambda_r - 2r\Lambda).$$

$$\gamma_{342} = \frac{1}{2}(\lambda_{342} + \lambda_{234} - \lambda_{423})$$

$$\lambda_{423} = 0$$

$$\lambda_{342} = -[\frac{l+a-i\overline{\rho}^*\csc\theta\cot\theta}{4\rho^2\overline{\rho}(\overline{\rho}^{*2})}\Lambda + \frac{i\csc\theta}{4\rho^2\overline{\rho}(\overline{\rho}^{*2})}\Lambda + \frac{i(l+a)\sin\theta}{4\rho^2\overline{\rho}(\overline{\rho}^{*2})}\Omega\varpi]$$

$$\lambda_{234} = \frac{1}{4\rho^2\overline{\rho}^*(\overline{\rho})^2}[i\rho(l+a)\cos\theta - ((l+a)\sin\theta)^2] - \frac{i(l+a)\sin\theta}{4\rho^2\overline{\rho}^*(\overline{\rho})^2}\Omega\varpi$$

$$\gamma_{342} = \frac{1}{2}\{\frac{1}{4\rho^2\overline{\rho}^*(\overline{\rho})^2}[i\rho(l+a)\cos\theta - ((l+a)\sin\theta)^2] - \frac{i(l+a)\sin\theta}{4\rho^2\overline{\rho}^*(\overline{\rho})^2}\Omega\varpi -$$

$$[\frac{l+a-i\overline{\rho}^*\csc\theta\cot\theta}{4\rho^2\overline{\rho}(\overline{\rho}^{*2})}\Lambda + \frac{i\csc\theta}{4\rho^2\overline{\rho}(\overline{\rho}^{*2})}\Lambda + \frac{i(l+a)\sin\theta}{4\rho^2\overline{\rho}(\overline{\rho}^{*2})}\Omega\varpi]\}$$

$$4\gamma = -\frac{\Omega \vartheta}{2\rho^6}(\rho^2 \Lambda_r - 2r\Lambda) + \frac{1}{2}\{\frac{1}{4\rho^2\overline{\rho}^*(\overline{\rho})^2}[i\rho(l+a)\cos\theta - ((l+a)\sin\theta)^2] - \frac{i(l+a)\sin\theta}{4\rho^2\overline{\rho}^*(\overline{\rho})^2}\Omega\varpi -$$

$$[\frac{l+a-i\overline{\rho}^*\csc\theta\cot\theta}{4\rho^2\overline{\rho}(\overline{\rho}^{*2})}\Lambda + \frac{i\csc\theta}{4\rho^2\overline{\rho}(\overline{\rho}^{*2})}\Lambda + \frac{i(l+a)\sin\theta}{4\rho^2\overline{\rho}(\overline{\rho}^{*2})}\Omega\varpi]\}$$

.

(30)

$$\alpha = \frac{1}{2}(\gamma_{214} + \gamma_{344})$$

$$\gamma_{214} = \frac{1}{2}(\lambda_{214} + \lambda_{421} - \lambda_{142})$$

$$\lambda_{214} = \frac{1}{2\sqrt{2}\rho^2 \bar{\rho}^* \Lambda^2}[(\Omega\varpi)_r \Lambda - \Omega\varpi\Lambda_r]$$

$$\lambda_{421} = -\frac{1}{2\sqrt{2}\rho^4 \bar{\rho}^*}\{[(\Omega\varpi)_r \rho^2 - 2r\Omega\varpi] - i(l+a)\sin\theta(\Lambda_r \rho^2 - 2r\Lambda)\}$$

$$-\lambda_{142} = -\frac{i\csc\theta}{\sqrt{2}(\rho\bar{\rho}^*)^2}\Omega\varpi$$

$$\gamma_{214} = \frac{1}{2}\{\frac{1}{2\sqrt{2}\rho^2 \bar{\rho}^* \Lambda^2}[(\Omega\varpi)_r \Lambda - \Omega\varpi\Lambda_r] - \frac{1}{2\sqrt{2}\rho^4 \bar{\rho}^*}\{[(\Omega\varpi)_r \rho^2 - 2r\Omega\varpi] -$$

$$i(l+a)\sin\theta(\Lambda_r \rho^2 - 2r\Lambda)\} - \frac{i\csc\theta}{\sqrt{2}(\rho\bar{\rho}^*)^2}\Omega\varpi\}$$

$$\gamma_{344} = \frac{1}{2}(\lambda_{344} + \lambda_{434} - \lambda_{443})$$

$$\lambda_{434} = 0$$

$$\lambda_{344} = \frac{i\csc\theta}{\sqrt{2}(\bar{\rho}^*)^3 \bar{\rho}}$$

$$\gamma_{344} = \frac{i\csc\theta}{\sqrt{2}(\bar{\rho}^*)^3 \bar{\rho}} \qquad (31)$$

$$\alpha = \frac{1}{4}\{\frac{1}{2\sqrt{2}\rho^2 \bar{\rho}^* \Lambda^2}[(\Omega\varpi)_r \Lambda - \Omega\varpi\Lambda_r] - \frac{1}{2\sqrt{2}\rho^4 \bar{\rho}^*}\{[(\Omega\varpi)_r \rho^2 - 2r\Omega\varpi] -$$

$$i(l+a)\sin\theta(\Lambda_r \rho^2 - 2r\Lambda)\} - \frac{i\csc\theta}{\sqrt{2}(\rho\bar{\rho}^*)^2}\Omega\varpi\} + \frac{i\csc\theta}{2\sqrt{2}(\bar{\rho}^*)^3 \bar{\rho}}$$

$$\beta = \frac{1}{2}(\gamma_{213} + \gamma_{343})$$

$$\gamma_{213} = \frac{1}{2}(\lambda_{213} + \lambda_{321} - \lambda_{132})$$

$$\lambda_{213} = -\frac{1}{2\sqrt{2}\bar{\rho}\rho^2}[(\Omega\varpi)_r - \frac{\Omega\varpi}{\Lambda}\Lambda_r]$$

$$\lambda_{321} = \frac{(l+a)^2 \sin\theta\cos\theta}{\sqrt{2}\rho^4\bar{\rho}}\Omega\vartheta - \frac{\rho^2\Lambda_r - 2r\Lambda}{2\sqrt{2}\rho^4\bar{\rho}}(i(l+a)\sin\theta) - \frac{\Omega\varpi}{2\sqrt{2}\rho^4\bar{\rho}}$$

$$-\lambda_{132} = -\frac{i\csc\theta}{\sqrt{2}(\rho\bar{\rho})^2}$$

$$\gamma_{213} = \frac{1}{2}\{-\frac{1}{2\sqrt{2}\bar{\rho}\rho^2}[(\Omega\varpi)_r - \frac{\Omega\varpi}{\Lambda}\Lambda_r] + \frac{(l+a)^2\sin\theta\cos\theta}{\sqrt{2}\rho^4\bar{\rho}}\Omega\vartheta -$$

$$\frac{\rho^2\Lambda_r - 2r\Lambda}{2\sqrt{2}\rho^4\bar{\rho}}(i(l+a)\sin\theta) - \frac{\Omega\varpi}{2\sqrt{2}\rho^4\bar{\rho}} - \frac{i\csc\theta}{\sqrt{2}(\rho\bar{\rho})^2}\}$$

$$\gamma_{343} = \frac{1}{2}(\lambda_{343} + \lambda_{334} - \lambda_{433})$$

$$\lambda_{343} = 0$$

$$\lambda_{334} = \frac{\csc\theta}{\sqrt{2}(G^2 + H^2)}(H - iG)$$

$$\gamma_{343} = \frac{\csc\theta}{\sqrt{2}(G^2 + H^2)}(H - iG) \quad , \quad (32)$$

$$\beta = \frac{1}{4}\{-\frac{1}{2\sqrt{2}\bar{\rho}\rho^2}[(\Omega\varpi)_r - \frac{\Omega\varpi}{\Lambda}\Lambda_r] + \frac{(l+a)^2\sin\theta\cos\theta}{\sqrt{2}\rho^4\bar{\rho}}\Omega\vartheta -$$

$$\frac{\rho^2\Lambda_r - 2r\Lambda}{2\sqrt{2}\rho^4\bar{\rho}}(i(l+a)\sin\theta) - \frac{\Omega\varpi}{2\sqrt{2}\rho^4\bar{\rho}} - \frac{i\csc\theta}{\sqrt{2}(\rho\bar{\rho})^2}\} + \frac{\csc\theta}{2\sqrt{2}(G^2 + H^2)}(H - iG)$$

where,

$$G = \{r^4 - 3r^2(l+a)^2\cos^2\theta[(l+a)^3\cos^3\theta - 3r^2(l+a)\cos\theta]\}$$
$$H = \{r(l+a)^3\cos^3\theta - 3r^3(l+a)\cos\theta - 3r(l+a)^3\cos^3\theta\} \quad . \quad (32a)$$

All the above pertain to the non-vanishing spin coefficients. We must note that in the above evaluations we have seen that $\varepsilon$ is not equal to zero, which suggests that the null geodesics so chosen for this metric are not affinely parameterized.

## VI. Conclusions and Discussions

Thus, with the aid of a null tetrad and the tool called N-P formalism, we were able to comprehend the physical nature of the G-P space time. We have observed that the chosen null tetrad for the G-P metric does indeed satisfy the conditions of shear-free, congruent null geodesics. It is not affinely parameterized which was also seen. The physical meanings of the various parameters that are involved have been assumed according to the G-P metric. The Maxwell field components $e$ and $g$ are related to the electric parameters and the magnetic charges of the sources. The twist of the principal null congruences which is proportional to $\omega$ is related to the Kerr rotation parameter $a$ and the NUT parameter $l$. The optical significance is also related to all these parameters.

## VII. Acknowledgements

This work was supported in grant from KVPY [Kishore Vaigyanik Protsahan Yojna], Dept. of Science, Govt. of India. I would like to thank Prof. Naresh Dadhich, Director, IUCAA [Inter University Center for Astronomy and Astrophysics] for his much needed support during this work. I have definitely been awarded with many valuable suggestions from Prof. Jerry B. Griffiths, Loughborough University. Finally, I would like to thank IUCAA for the congenial environment provided during the course of this work.

The original papers in which Petrov Classification appeared is described is,